\documentclass[11pt]{article}
\usepackage{epsfig,latexsym,amsmath,emp}
\makeatletter
\def\@normalsize{\@setsize\normalsize{12pt}\xpt\@xpt
\abovedisplayskip 10pt plus2pt minus5pt\belowdisplayskip \abovedisplayskip
\abovedisplayshortskip \z@ plus3pt\belowdisplayshortskip 6pt plus3pt
minus3pt\let\@listi\@listI}

\def\subsize{\@setsize\subsize{12pt}\xipt\@xipt}
\newcommand{\excise}[1]{}
\newcommand{\nix}[1]{}

\newcommand{\ket}[1]{|#1\rangle}
\newcommand{\proof}{\textit{Proof. }} 
\newcommand{\scal}[2]{\langle #1|#2\rangle}
\newcommand{\qed}{\mbox{$\Box$}}
\newcommand{\tr}{\mathop{\textup{tr}}}
\newcommand{\diag}{\textup{diag}}
\newcommand{\antidiag}{\textup{antidiag}}

\newcommand{\mat}[4]{\left(\begin{array}{cc}#1 & #2\\ #3 & #4\end{array}\right)}
\newcommand{\rmat}[4]{\left(\begin{array}{rr}#1 & #2\\ #3 & #4\end{array}\right)}

\newtheorem{theorem}{Theorem}
\newtheorem{lemma}[theorem]{Lemma}
\newtheorem{corollary}[theorem]{Corollary}

\newtheorem{proposition}[theorem]{Proposition}
\newtheorem{remark}[theorem]{Remark}

\newcommand{\C}{\mathbf{C}}
\newcommand{\R}{\mathbf{R}}

\makeatother

\begin{document}
\date{}

\title{\vspace{-1cm}\Large\textbf{Optimal Realizations of Controlled Unitary Gates}} 

\author{\normalsize
  {Guang Song and Andreas Klappenecker}\\
  {Department of Computer Science}\\
  {Texas A\&M University} \\
  {College Station, TX~~77843-3112}\\
  {\{\texttt{gsong,klappi}\}\texttt{@cs.tamu.edu}}}

\maketitle

\subsection*{\centering Abstract}

{\small \em The controlled-not gate and the single qubit gates are
considered elementary gates in quantum computing.  It is natural to
ask how many such elementary gates are needed to implement more
elaborate gates or circuits. Recall that a controlled-$U$ gate can be
realized with two controlled-not gates and four single qubit gates. We
prove that this implementation is optimal if and only if the matrix
$U$ satisfies the conditions $\tr U\neq 0$, $\tr(UX)\neq 0$, and $\det
U\neq 1$. We also derive optimal implementations in the remaining non-generic cases. 
}

\def\thefootnote{\arabic{footnote}}
\setcounter{footnote}{0}

\section{Introduction}\label{sec:intro}\enlargethispage{4mm}
It was shown in the seminal paper~\cite{bbc-egqc-95} that any unitary
$2^n\times 2^n$ matrix $M$ can be realized on a quantum computer with
$n$ quantum bits by a finite sequence of controlled-not and single
qubit gates. We will refer to controlled-not and single qubit gates as
elementary gates. It is natural to ask how many elementary gates are
necessary and sufficient to realize a given unitary matrix $M$.
Answering such questions is a notoriously difficult task. 

It was shown in~\cite{bbc-egqc-95}
that a controlled unitary operation can be realized with at most six
elementary gates, that is, given a unitary $2\times 2$ matrix $U$,
there exist unitary matrices $A, B, C,$ and $E$ such that 
\empprelude{prologues := 0;}
\begin{empfile}
\begin{equation}
\begin{emp}(50,50)
  setunit 2mm;
  qubits(2);
  gate(icnd 1, gpos 0, btex $U$ etex);
  label(btex $=$ etex,(QCxcoord+1/2QCstepsize, QCycoord[0]+3mm));
  QCxcoord := QCxcoord + QCstepsize;
  QCstepsize := QCstepsize-4mm;
  wires(2mm);
  gate(gpos 1, btex $E$ etex, 0, btex $A$ etex);
  cnot(icnd 1, gpos 0);
  gate(gpos 0, btex $B$ etex);
  cnot(icnd 1, gpos 0);
  gate(gpos 0, btex $C$ etex);
  wires(2mm);
\end{emp}
\end{equation}
Our main result shows that this implementation is optimal:
\medskip

\noindent\textbf{Theorem A}
\textit{Suppose that $U$ is a unitary $2\times 2$ matrix satisfying 
$\det U\neq 1$, $\tr U\neq 0$, and\/ $\tr UX\neq 0$. Then six elementary
gates are necessary and sufficient to implement a controlled-$U$ gate.
}
\medskip

\noindent\textit{Notations.} 
We use the following abbreviations throughout this paper:
$$ 
H = \frac{1}{\sqrt{2}}\rmat{1}{1}{1}{-1},\quad X =
\mat{0}{1}{1}{0},\quad Z = \rmat{1}{0}{0}{-1}.$$ We denote by $\C$ the
field of complex numbers, and by $\R$ the field of real numbers. 
We will say that a unitary matrix $U$ is generic if and only if the
conditions $\det U\neq 1$, $\tr U\neq 0$, and $\tr UX\neq 0$ are
satisfied.  Notice that the inverse $U^\dagger$ of a generic unitary
matrix $U$ is again generic.
\medskip

Theorem A gives a sharp lower bound on the number of elementary gates
that are needed to implement a generic controlled-$U$ operation. The
non-generic case is discussed in Theorem B in
Section~\ref{sec:nongeneric}.

\section{Proof of Theorem A}
We will show that any implementation of a generic controlled-$U$
operation requires at least six elementary gates. We classify the
possible implementation in terms of the number of controlled-not
operations used. We will use entanglement properties to rule out
various potential implementations. The following simple fact will turn out
to be particularly helpful:

\begin{lemma}\label{lemma:entanglement}
Let $\ket{\psi}$, $\ket{\phi}$ be nonzero elements of $\C^2$.  
The input\/
$\ket{\psi}\otimes \ket{\phi}$ to a controlled-$U$ gate will produce
an entangled output state if and only if\/ $\ket{\phi}$ is not an eigenvector of\/ $U$ and\/ $\ket{\psi}=a\ket{0}+b\ket{1}$ 
with\/ $a,b\neq 0$. 
\end{lemma}
\proof The input $\ket{\psi}\otimes \ket{\phi}= (a\ket{0}+b\ket{1})\otimes \ket{\phi}$ to the controlled-$U$ gate produces the result 
$$ \ket{r_{out}} = a\ket{0}\otimes \ket{\phi}+b\ket{1}\otimes U\ket{\phi}$$ 
Denote by $\ket{\phi^\perp}$ a nonzero vector in $\C^2$ 
satisfying $\scal{\phi^\perp}{\phi}=0$. 
Consequently, $U\ket{\phi} = c\ket{\phi}+d\ket{\phi^\perp}$ with 
$c,d\in \C$, and $d\neq 0$. Therefore, the output state 
$\ket{r_{out}}$ can be expressed in the form
\begin{equation}\label{eq:entangledstate} 
\ket{r_{out}} = a\,\ket{0}\otimes \ket{\phi}
+bc\,\ket{1}\otimes\ket{\phi}
+bd\,\ket{1}\otimes\ket{\phi^\perp}
\end{equation}
with $a,b,d\neq 0$. Seeking a contradiction, we assume that 
$\ket{r_{out}}$ is not an entangled state. This would mean that there exist complex coefficients $\alpha, \beta, \gamma, \delta$ such that 
$$
\begin{array}{lcl}
\ket{r_{out}} &=&
(\alpha\ket{0}+\beta\ket{1})\otimes 
(\gamma\ket{\phi}+\delta\ket{\phi^\perp})\\[1ex]
&=&
\alpha\gamma\ket{0}\otimes \ket{\phi} + \alpha\delta\ket{0}\otimes 
\ket{\phi^\perp}+\beta\gamma\ket{1}\otimes \ket{\phi}+
\beta\delta\ket{1}\otimes \ket{\phi^\perp}.
\end{array}
$$ 
Comparing coefficients with (\ref{eq:entangledstate}) shows that
$\alpha\delta=0$, hence $\alpha$ or $\delta$ has to be zero. Either
choice leads to a contradiction.  

On the other hand, if $\ket{\psi}$ is a multiple of $\ket{0}$ or of
$\ket{1}$ or if $\ket{\phi}$ is an eigenvector of $U$, then it follows
from the definitions that the output of the controlled-$U$ gate will
not be entangled.~\qed

\begin{corollary}\label{corollary}
Assume that $\ket{\phi}$ is an eigenvector of $U$ with eigenvalue $\lambda_\phi$, 
and a state $\ket{\psi}\in \C^2$.  If we
input $\ket{\psi}\otimes \ket{\phi}$ to the controlled-$U$ gate, then
the output is of the form $\diag(1,\lambda_\phi)\ket{\psi}\otimes \ket{\phi}$. 
In particular, the output is not entangled. 
\end{corollary}

\goodbreak Another simple consequence of this lemma is that a controlled-$U$
gate is able to produce an entangled output state from some input
state $\ket{\psi}\otimes \ket{\phi}$, as long as $U$ is not a multiple
of the identity matrix. Any matrix $U$ with $\tr(UX)\neq 0$ satifies
this condition. In particular, we need at least one controlled-not
gate to implement a controlled-$U$ gate with $\tr(UX)\neq 0$.

\paragraph{One Controlled-Not Gate.} 
We consider now possible implementations of a generic controlled-$U$
gate with only one controlled-not gate and some single qubit
gates. Recall that it is possible to switch the control and the target
qubit of a controlled-not gate by conjugation with Hadamard matrices
$H$:
\begin{equation}\label{circ:flip}
\begin{emp}(50,50)
  setunit 2mm; qubits(2);
  
  cnot(icnd 0, gpos 1);
  label(btex $=$ etex,(QCxcoord+1/2QCstepsize, QCycoord[0]+3mm));
  QCxcoord := QCxcoord + QCstepsize;
  QCstepsize := QCstepsize-4mm;
  wires(2mm);
  gate(gpos 1, btex $H$ etex, 0, btex $H$ etex);
  cnot(icnd 1, gpos 0);
  gate(gpos 1, btex $H$ etex, 0, btex $H$ etex);
  wires(2mm);
\end{emp}
\end{equation}
Therefore, we can assume without loss of generality that the controlled-$U$ gate is expressed in the following form:
\begin{equation}\label{circ:single}
\begin{emp}(50,50)
  setunit 2mm; qubits(2);
  
  gate(icnd 1, gpos 0, btex $U$ etex);
  label(btex $=$ etex,(QCxcoord+1/2QCstepsize, QCycoord[0]+3mm));
  QCxcoord := QCxcoord + QCstepsize;
  QCstepsize := QCstepsize-4mm;
  wires(2mm);
  gate(gpos 1, btex $A_1$ etex, 0, btex $B_1$ etex);
  cnot(icnd 1, gpos 0);
  gate(gpos 1, btex $A_2$ etex, 0, btex $B_2$ etex);
  wires(2mm);
\end{emp}
\end{equation}

\begin{lemma}\label{lemma:single}
The unitary matrices $A_1, A_2$ used in the single qubit gates in
(\ref{circ:single}) have to be both diagonal or both antidiagonal.
\end{lemma}
\proof Suppose that we input $\ket{i} \otimes \ket{\alpha}$, where
$i=0,1$ and $\ket{\alpha}$ is some arbitrary vector in $\C^2$ such
that $B_1\ket{\alpha}$ is not an eigenvector of $X$.
Lemma~\ref{lemma:entanglement} shows that the output of the
controlled-$U$ gate is not entangled when provided with such an input
state, since the most significant qubit is not in superposition. 
Notice that the circuit on the right hand side in
(\ref{circ:single}) will produce an entangled output unless $A_1$ is
diagonal or antidiagonal. It follows that $A_1$ is of the desired
form. The same argument applied to the inverse circuits proves that
$A_2$ has to be diagonal or antidiagonal.  It is clear that $A_1$ and
$A_2$ are either both diagonal or both antidiagonal, because
$\ket{00}$ has to be an eigenstate of the
circuit~(\ref{circ:single}).~\qed
\medskip

We can assume that $A_1$ and $A_2$ are diagonal. Indeed, if the
$A_i$'s are antidiagonal, then we can replace the controlled-not gate
in~(\ref{circ:single}) by 
\begin{equation}\label{circ:cnot}
\begin{emp}(50,50)
  setunit 2mm; qubits(2);
  
  cnot(icnd 1, gpos 0);
  label(btex $=$ etex,(QCxcoord+1/2QCstepsize, QCycoord[0]+3mm));
  QCxcoord := QCxcoord + QCstepsize;
  QCstepsize := QCstepsize-4mm;
  wires(2mm);
  gate(gpos 1, btex $X$ etex);
  cnot(icnd 1, gpos 0);
  gate(gpos 0, btex $X$ etex, 1, btex $X$ etex);
  wires(2mm);
\end{emp}
\end{equation}
Here we used the fact that $XA_1$ and $A_2X$ will be both diagonal, when $A_1, A_2$ are antidiagonal.

\begin{lemma}\label{lemma:onecontrol}
If $\tr U\neq 0$, then the circuit (\ref{circ:single}) cannot implement a 
controlled-$U$ operation.
\end{lemma}
\proof The preceding discussion shows that $A_1$ and $A_2$ are both
diagonal or antidiagonal, and we may assume that $A_1$ and $A_2$ are
diagonal. Thus, there exist real numbers $\vartheta_0$ and
$\vartheta_1$ such that $A_2A_1\ket{0}=e^{i\vartheta_0}\ket{0}$ and
$A_2A_1\ket{1}=e^{i\vartheta_1}\ket{1}$.  It follows that
$B_2B_1=e^{-i\vartheta_0}I$ and $B_2XB_1=e^{-i\vartheta_1}U$.
Consequently, $B_1^\dagger XB_1=e^{i(\vartheta_0-\vartheta_1)}U$,
which implies $\tr U = 0$.  Therefore, it is in general not possible
to implement a controlled-$U$ operation with one controlled-not gate
and several single qubit gates.~\qed
 
\paragraph{Two Controlled-Not Gates.}
Assume that we have now two controlled-not gates and several single
qubit gates at our disposal. This allows to express the controlled-$U$ gate in the form
\begin{equation}\label{circ:double}
\vbox{\hbox{
\begin{emp}(50,50)
  setunit 2mm;
  qubits(2);
  
  gate(icnd 1, gpos 0, btex $U$ etex);
  label(btex $=$ etex,(QCxcoord+1/2QCstepsize, QCycoord[0]+3mm));
  QCxcoord := QCxcoord + QCstepsize;
  QCstepsize := QCstepsize-4mm;
  wires(2mm);
  gate(gpos 1, btex $A_1$ etex, 0, btex $B_1$ etex);
  cnot(icnd 1, gpos 0);
  gate(gpos 1, btex $A_2$ etex, 0, btex $B_2$ etex);
  cnot(icnd 1, gpos 0);
  gate(gpos 1, btex $A_3$ etex, 0, btex $B_3$ etex);

  wires(2mm);
\end{emp}
}}
\end{equation}
In fact, any implementation of a controlled-$U$ gate with two
controlled-not gates can be reduced to this form.  Indeed, it is
possible to swap the control and target qubits of a controlled-not
gate by conjugation with Hadamard gates, as we have seen in our
discussion of the previous case. We will see what kind of properties
have to be satisfied by the matrices $A_i$ and $B_i$. The following
Lemmas will prepare us to prove Proposition~\ref{proposition}.

We say that a unitary $2\times 2$ matrix $V$ is sparse if and only if
$V$ is diagonal or antidiagonal.

\begin{lemma}\label{lemma:sparse} 
If the matrix $A_1$ in (\ref{circ:double}) is sparse, 
then $A_2, A_3$  are sparse as well. 
\end{lemma}
\proof Suppose that $A_1$ is diagonal. Choose a state $\ket{0}\otimes
B_1^\dagger B_2^\dagger\ket{\psi}$ as input, where 
$\ket{\psi}$ is not an eigenstate of $X$. Notice that a
controlled-$U$ operation leaves the input $\ket{0}\otimes \ket{\psi}$
invariant. Consider now the evolution of the input state through the
circuit on the right hand side of (\ref{circ:double}).  Because $A_1$
is diagonal, the resulting state after the first controlled-not
operation is $\alpha\ket{0}\otimes B_2^\dagger \ket{\psi} $, where
$\alpha$ is a scalar phase factor.  Applying the single qubit
operations $A_2$ and $B_2$ yields $\alpha A_2\ket{0}\otimes
\ket{\psi}$. Lemma~\ref{lemma:entanglement} shows that the output
after the second controlled-not operation will be entangled, unless
$A_2$ is sparse. Therefore, $A_2$ has to be sparse.  Thus, the state
after the second controlled-not is of the form $\beta\ket{i}\otimes
\ket{\psi'}$, where $\beta$ is some phase factor, $i=0,1$, and
$\ket{\psi'}$ is some element of $\C^2$. Hence $A_3$ has to map
$\ket{i}$ to $\ket{0}$, up to a phase factor, i.e., $A_3$ has to be
sparse.

If $A_1$ is antidiagonal, then we can use the identity
(\ref{circ:cnot}) to replace the antidiagonal matrix $A_1$ by the 
diagonal matrix $XA_1$, which allows us to conclude that $A_2X$, 
hence $A_2$, and $A_3$ are sparse.~\qed

\begin{lemma}\label{lemma:notsparse}
Suppose that $U$ is not a multiple of the identity matrix. 
If $A_1$ in the circuit (\ref{circ:double}) is 
not sparse, then $A_2, A_3$ are not sparse either.
\end{lemma}
\proof Assume that the input state is of the form $A_1^\dagger\ket{0}
\otimes \ket{\psi}$, where $\ket{\psi}$ is not an eigenvector of $U$.
Since $A_1$ is not sparse, $A_1^\dagger \ket{0}=a\ket{0}+b\ket{1}$
with $a,b\neq 0$. Therefore, the input $A_1^\dagger\ket{0} \otimes
\ket{\psi}$ to the controlled-$U$ operation will yield an entangled
output state, according to Lemma~\ref{lemma:entanglement}. 

On the other hand, consider the right hand side of
(\ref{circ:double}).  The input state $A_1^\dagger\ket{0}\otimes
\ket{\psi}$ produces after the first controlled-not gate a state of
the form $\ket{0}\otimes B_1\ket{\psi}$. The input to the second
controlled-not gate is then $A_2\ket{0}\otimes
B_2B_1\ket{\psi}$. Lemma~\ref{lemma:entanglement} shows that the output
of the second controlled-not operation cannot be entangled, unless
$A_2$ is not sparse.  Therefore, $A_2$ is not sparse. 
However, $A_3$ cannot be sparse either, because
this would imply that $A_2$ and $A_1$ are sparse, as can be seen by
applying Lemma~\ref{lemma:sparse} to the inverse circuit.~\qed
\medskip

\begin{lemma}\label{lemma:Asparse}
Let\/ $U$ be a unitary $2\times 2$ matrix.  
Assume that $A_1, A_2, A_3$ in (\ref{circ:double}) are
sparse. If  $\tr U\neq 0$, then $B_2\neq H$. If $\tr(UX)\neq 0$,  
then none of the matrices $B_1, B_2, B_3$ can be equal to an
identity matrix, and $B_1, B_3$ cannot both be equal to
$H$.
\end{lemma}
\proof Comparing the result of the inputs $\ket{0}\otimes \ket{\psi}$ and 
 $\ket{1}\otimes \ket{\psi}$ on the left and right hand side of (\ref{circ:double}) yields
\begin{align}
\label{eq:contrA} e^{i\theta_0} I &= B_3X^{k}B_2X^{\ell}B_1,\\
\label{eq:contrB} e^{i\theta_1} U &= B_3X^{1-k}B_2X^{1-\ell}B_1,
\end{align}
for some $k,\ell\in \{0,1\}.$ 
Notice that equation (\ref{eq:contrA}) implies $B_2=e^{i\theta_0}X^k B_3^\dagger B_1^\dagger X^\ell$. Substituting $B_2$ in (\ref{eq:contrB}) yields 
\begin{equation}\label{eq:U}
U = e^{i(\theta_0-\theta_1)}B_3X B_3^\dagger B_1^\dagger X B_1.   
\end{equation}
\goodbreak

\noindent\textit{Step 1.}
We show that $B_i\neq I$ for $i=1,2,3$:

\smallskip
\noindent i) Suppose that $B_1=I$. Equation~(\ref{eq:U}) implies  
$\tr(UX)=0$, contradicting our assumptions. 

\smallskip
\noindent ii) Suppose that $B_2=I$. Equations (\ref{eq:contrA}) and (\ref{eq:contrB}) then imply $U=e^{i(\theta_0-\theta_1)}I$, 
thus $\tr(UX)=0$, which contradicts our assumptions. 

\smallskip\noindent iii) Suppose that $B_3=I$. 
Equation~(\ref{eq:U}) implies $\tr(XU)=0$, whence $\tr(UX)=0$.
This contradicts our assumptions.
\medskip\goodbreak

\noindent\textit{Step 2.} We show that $B_2\neq H$. Seeking a
contradiction, we suppose that $B_2=H$. From (\ref{eq:contrA}) and
(\ref{eq:contrB}), $U=e^{i(\theta_0-\theta_1)}
B_3X^{1-k}ZX^kB_3^\dagger$, which implies $\tr(U)=0$. 
Contradiction. 

\medskip
\noindent\textit{Step 3.} 
The case $B_1=B_3=H$ immediately leads to a contradiction, 
because (\ref{eq:U}) would imply 
$U=e^{i(\theta_0-\theta_1)}I$, and thus $\tr(UX)=0$.~\qed

\begin{lemma}\label{lemma:det}
If\/ $\det U\neq 1$, then at least one of the matrices $A_i$ in (\ref{circ:double}) is not equal to the identity matrix. 
\end{lemma}
\proof Seeking a contradiction, we assume that $A_1, A_2, A_3$ are
identity matrices. It follows at once that $B_3B_2B_1=e^{i\phi}I$, and
$B_3XB_2XB_1=e^{i\phi}U$, hence $\det U=1$. It
follows that one of the matrices $A_i$ has to differ from the identity
matrix.~\qed

\begin{lemma}\label{lemma:eigenvector}
If the matrices $A_1, A_2,
A_3$ in (\ref{circ:double}) are not sparse, then $B_2\neq H$ and 
$B_1^\dagger \ket{\omega_0}$ and $B_1^\dagger \ket{\omega_1}$ are 
eigenvectors of\/ $U$, where 
$$\ket{\omega_0}=(\ket{0}+\ket{1})/\sqrt{2}\quad \mbox{and}\quad
\ket{\omega_1}=(\ket{0}-\ket{1})/\sqrt{2}$$
are eigenvectors of $X$. 
\end{lemma}
\proof Consider the input $\ket{i}\otimes B_1^\dagger\ket{\omega_k}$,
with $i,k=0,1$. Note that the controlled-$U$ operation does not
produce an entangled output state when provided with such an input.
On the other hand, consider the evolution of these states in the
circuit~(\ref{circ:double}). The first two single qubit operations
yield the state $A_1\ket{i}\otimes \ket{\omega_k}$. The controlled-not
operation produces the state $Z^kA_1\ket{i}\otimes \ket{\omega_k}$,
where we have used the fact that $\ket{\omega_k}$ is an eigenvector of
$X$ with eigenvalue $(-1)^k$.  The result of the next two single qubit
operations is then $A_2Z^k A_1\ket{i}\otimes
B_2\ket{\omega_k}$. Notice that the matrix $A_2Z^kA_1$ cannot be
sparse, because this would imply that $A_3$ is sparse. In other words,
the input to the second controlled-not gate is a state of the form
$(a_i\ket{0}+b_i\ket{1})\otimes B_2\ket{\omega_k}$ with $a_i, b_i\neq
0$.  Since the circuit implements a controlled-$U$ operation, this
gate should not produce an entangled output state. Therefore,
$B_2\ket{\omega_k}$ has to be an eigenvector of $X$.  However, this
means that the input $\ket{\psi}\otimes B_1^\dagger \ket{\omega_k}$ to
(\ref{circ:double}) does not get entangled for arbitrary states
$\ket{\psi}$.  Consequently, $B_1^\dagger \ket{\omega_k}$ has to be an
eigenvector of $U$ by Lemma~\ref{lemma:entanglement}.

The previous discussion showed that $B_2\ket{\omega_k}$, $k=0,1$, has
to be an eigenvector of $X$, i.e., is mapped to a multiple of
$\ket{\omega_\ell}$ for $\ell=0,1$. In particular, 
$B_2$ cannot be the Hadamard matrix $H$.~\qed

\begin{lemma}\label{lemma:double}
If the matrices $A_1, A_2,A_3$ in the circuit (\ref{circ:double}) are
all nonsparse, then either $A_3A_2A_1$ or $A_3ZA_2A_1$ is a diagonal matrix.
In particular, it is not possible that 
$A_i=A_{i+1}=H$ for $i=1,2$. 
\end{lemma}
\proof Recall that $B_1^\dagger \ket{\omega_0}$ is an eigenstate of
$U$, where $\ket{\omega_0}=(\ket{0}+\ket{1})/\sqrt{2}$, as is shown in
Lemma~\ref{lemma:eigenvector}. 
The circuit on the right hand side of
(\ref{circ:double}) maps the input $\ket{0}\otimes B_1^\dagger
\ket{\omega_0}$ to $A_3Z^\ell A_2A_1\ket{0}\otimes
B_3B_2\ket{\omega_0}$, where $\ell=0$ if $B_2$ maps $\ket{\omega_0}$ to a multiple of itself, and $\ell=1$ otherwise. 
Comparing this state with the supposed output state $\ket{0}\otimes
B_1^\dagger\ket{\omega_0}$ shows that $A_3Z^\ell A_2A_1\ket{0}$ coincides, up
to a phase factor, with $\ket{0}$. Hence $A_3Z^\ell A_2A_1$ has to be a
diagonal unitary matrix. The second statement is obvious.~\qed

\begin{lemma}\label{lemma:Anotsparse}
Let $U$ be a unitary $2\times 2$ matrix with $\tr(UX)\neq 0$. 
If $A_1,
A_2, A_3$ in (\ref{circ:double}) are not sparse, then $B_1,B_2,B_3\neq
H$, and at least one of the matrices $B_1, B_2, B_3$ differs from the
identity matrix.
\end{lemma}
\proof Let $\ket{\omega_0}=(\ket{0}+\ket{1})/\sqrt{2}$ and
$\ket{\omega_1}=(\ket{0}-\ket{1})/\sqrt{2}$. Lemma~\ref{lemma:eigenvector}
showed that $B_1^\dagger \ket{\omega_0}$ and $B_1^\dagger
\ket{\omega_1}$ are eigenvectors of $U$, that is,
$UB_1^\dagger\ket{\omega_k} = \alpha_k B_1^\dagger\ket{\omega_k}$ for
$k=0,1$. Hence $B_1UB_1^\dagger = H\,\diag(\alpha_0,\alpha_1)H.$ The
choice $B_1=H$ would force $U$ to be diagonal, which would contradict
$\tr(UX)\neq 0$. The same argument applied to the inverse circuit
proves $B_3\neq H$. We already know that $B_2\neq H$ by
Lemma~\ref{lemma:eigenvector}.

Seeking a contradiction, we assume that $B_1=B_2=B_3=I$. 
A potential implementation of the controlled-$U$ gate is given by:
\begin{equation}
\begin{emp}(50,50)
  setunit 2mm;
  qubits(2);

  gate(icnd 1, gpos 0, btex $U$ etex);
  QCstepsize := QCstepsize - 10;

  label(btex $=$ etex,(QCxcoord+1/2QCstepsize, QCycoord[0]+3mm));
  QCxcoord := QCxcoord + QCstepsize;

  wires(2mm);
  gate(gpos 1, btex $A_1$ etex);
  cnot(icnd 1, gpos 0);
  gate(gpos 1, btex $A_2$ etex);
  cnot(icnd 1, gpos 0);
  gate(gpos 1, btex $A_3$ etex);
  wires(2mm);
\end{emp}
\end{equation}
The choice $B_1=I$ implies $U=B_1UB_1^\dagger =
H\,\diag(\alpha_0,\alpha_1)H$, according to our discussion above. This
special form of $U$ shows that $\ket{\omega_0}$ and $\ket{\omega_1}$
are eigenvectors of $U$ corresponding to the eigenvalues $\alpha_0$
and $\alpha_1$. If we input a state $\ket{\psi}\otimes
\ket{\omega_0}$,  then Corollary~\ref{corollary} shows that
$A_3A_2A_1=\diag(1,\alpha_0)$. Similarly, we obtain 
$A_3ZA_2ZA_1=\diag(1,\alpha_1)$, considering input states of the form  
$\ket{\psi}\otimes \ket{\omega_1}$.

The determinants of $A_3A_2A_1$ and $A_3ZA_2ZA_1$ are the same, hence
$\alpha_0$ has to coincide with $\alpha_1$. However, this implies that
$U$ is diagonal, because 
$U=H\diag(\alpha_0,\alpha_1)H=\diag(\alpha_0,\alpha_0)$, whence $\tr(UX)=0$. 
Therefore, at least one of 
the matrices $B_i$ has to differ from the identity matrix.~\qed

\medskip
\begin{lemma}\label{lemma:obstruction}
If\/ $\det U\neq 1$, then the circuit~(\ref{circ:upside}) cannot
implement a controlled-$U$ gate.
\begin{equation}\label{circ:upside}
\vbox{\hbox{
\begin{emp}(50,50)
  setunit 2mm;
  qubits(2);
  QCstepsize := QCstepsize - 10;
 
  wires(2mm);
  gate(gpos 0, btex $B_1$ etex);
  cnot(icnd 0, gpos 1);
  gate(gpos 1, btex $C$ etex);
  cnot(icnd 0, gpos 1);
  gate(gpos 0, btex $B_3$ etex);
  wires(2mm);
\end{emp}
}}
\end{equation}
\end{lemma}
\proof Transforming (\ref{circ:upside}) into the form
(\ref{circ:double}) yields $A_1=H$, $A_2=HCH$, and $A_3=H$.
Lemma~\ref{lemma:double} shows that $A_3Z^\ell
A_2A_1=\diag(\alpha_0,\alpha_1)$, with $\ell=0$. Therefore,
$C=\diag(\alpha_0,\alpha_1)$. A diagonal matrix $C$ satisfies
\begin{equation*}
\begin{emp}(50,50)
  setunit 2mm;
  qubits(2);
  QCstepsize := QCstepsize - 10;
 
  wires(2mm);
  cnot(icnd 0, gpos 1);
  gate(gpos 1, btex $C$ etex);
  cnot(icnd 0, gpos 1);
  wires(2mm);

  label(btex $=$ etex,(QCxcoord+1/2QCstepsize, QCycoord[0]+3mm));
  QCxcoord := QCxcoord + QCstepsize;
  
  wires(2mm);
  cnot(icnd 1, gpos 0);
  gate(gpos 0, btex $C$ etex);
  cnot(icnd 1, gpos 0);
  wires(2mm);

\end{emp}
\end{equation*}
It follows that the circuit (\ref{circ:upside}) can be written in the
form (\ref{circ:double}) with $A_i=I$ for $i=1,2,3$. This contradicts
Lemma~\ref{lemma:det}.~\qed

\begin{proposition}\label{proposition}
Suppose that $U$ is a unitary $2\times 2$ matrix satisfying $\tr U\neq
0$ and\/ $\tr(UX)\neq 0$. Any implementation of a controlled-$U$ gate
with two controlled-not gates and some single qubit gates needs at
least a total of six gates provided that\/ $\det U\neq 1$, and at least
a total of five gates otherwise.
\end{proposition}
\proof Suppose we are given a fixed control-$U$ gate.  The
implementations of a controlled-$U$ gate with two controlled not gates
and single qubit gates can be classified according to the positions of
the target qubits of the two controlled-not gates.  We will show that
any of the four implementation types will require at least six
elementary gates.
\medskip

\noindent\textit{Case 1.} Suppose that the target bit of both
controlled-not operations is the least significant bit, as shown in
(\ref{circ:double}).

Suppose that $A_1$ is sparse.  We know from Lemma~\ref{lemma:Asparse}
that none of the matrices $B_i$, $i=1,2,3$, 
can be an identity matrix, whence we
have a total of five or more gates.  If $\det U\neq 1$, then
Lemma~\ref{lemma:det} shows that at least one of the matrices $A_i$ is
not the identity matrix, giving an additional gate.

Suppose that $A_1$ is not sparse. Then $A_2$ and $A_3$ are not sparse
either, by Lemma~\ref{lemma:notsparse}. We know from
Lemma~\ref{lemma:Anotsparse} that 
at least one of the matrices $B_1, B_2, B_3$ is not an identity matrix, 
whence we have a total of at least six gates.

\medskip
\noindent\textit{Case 2.} Suppose that the first controlled-not gate
acts on the most significant bit and the second controlled-not gate
acts on the least significant bit. So the circuit is of the form:
\begin{equation}\label{circ:twisted}
\begin{emp}(50,50)
  setunit 2mm;
  qubits(2);
  QCstepsize := QCstepsize-4mm;
  
  wires(2mm);
  gate(gpos 1, btex $C_1$ etex, 0, btex $D_1$ etex);
  cnot(icnd 0, gpos 1);
  gate(gpos 1, btex $C_2$ etex, 0, btex $D_2$ etex);
  cnot(icnd 1, gpos 0);
  gate(gpos 1, btex $C_3$ etex, 0, btex $D_3$ etex);
  wires(2mm);
  label(btex $=$ etex,(QCxcoord+1/2QCstepsize, QCycoord[0]+3mm));
  QCxcoord := QCxcoord + QCstepsize;

  wires(2mm);
  gate(gpos 1, btex\small $H\!C_1$ etex, 0, btex \small $H\!D_1$ etex);
  cnot(icnd 1, gpos 0);
  gate(gpos 1, btex\small $C_2\!H$ etex, 0, btex \small $D_2\!H$ etex);
  cnot(icnd 1, gpos 0);
  gate(gpos 1, btex $C_3$ etex, 0, btex  $D_3$ etex);
  wires(2mm);
\end{emp}
\end{equation}
We use the circuit on the right hand to show that the circuit on the
left hand side cannot have less than six elementary gates.

Assume that $HC_1$ is sparse. Then $C_2H$ has to be sparse as well,
hence $C_1$ and $C_2$ cannot be identity
matrices. Lemma~\ref{lemma:Asparse} shows that $D_3\neq I$ and
$D_2H\neq H$, hence $D_2\neq I$. Thus we have at least six elementary
gates.

Assume that $HC_1$ is not sparse. Then $C_3$ cannot be sparse.  Either
$C_1$ or $C_2$ has to differ from the identity matrix, because
Lemma~\ref{lemma:double} shows that $HC_1$ and $C_2H$ cannot both be
equal to $H$. Lemma~\ref{lemma:Anotsparse} shows that $HD_1\neq H$ and
$D_2H\neq H$, hence $D_1, D_2\neq I$. Thus we have at least six
elementary gates.

\medskip\noindent\textit{Case 3.} Suppose that the first
controlled-not gate acts on the least significant bit and the second
controlled-not gate acts on the most significant bit.  The inverse
circuit cannot implement a controlled-$U^\dagger$ operation with less
than six elementary gates, because it is of the form discussed in Case
2.\medskip

\goodbreak
\noindent\textit{Case 4.}  Finally, suppose that the target qubit of 
both controlled-not gates is the most significant qubit. Thus, the circuit is 
of the form 
\begin{equation}\label{circ:top}
\begin{emp}(50,50)
  setunit 2mm;
  qubits(2);
  QCstepsize := QCstepsize-4mm;
  
  wires(2mm);
  gate(gpos 1, btex $C_1$ etex, 0, btex $D_1$ etex);
  cnot(icnd 0, gpos 1);
  gate(gpos 1, btex $C_2$ etex, 0, btex $D_2$ etex);
  cnot(icnd 0, gpos 1);
  gate(gpos 1, btex $C_3$ etex, 0, btex $D_3$ etex);
  wires(2mm);
  label(btex $=$ etex,(QCxcoord+1/2QCstepsize, QCycoord[0]+3mm));
  QCxcoord := QCxcoord + QCstepsize;

  wires(2mm);
  gate(gpos 1, btex\small $HC_1$ etex, 0, btex \small $HD_1$ etex);
  cnot(icnd 1, gpos 0);
  QCxsave := QCxcoord;
  circuit(1.2cm)(gpos 1,1, btex \small $HC_2H$ etex);
  QCxcoord := QCxsave;
  QCqubitnum := 1;
  circuit(1.2cm)(gpos 0,0, btex \small $HD_2H$ etex);
  QCqubitnum := 2;
  cnot(icnd 1, gpos 0);
  gate(gpos 1, btex \small $C_3H$ etex, 0, btex \small $D_3\!H$ etex);
  wires(2mm);
\end{emp}
\end{equation}
Assume that $HC_1$ is sparse, then $C_3H$ is sparse as well, thus
$C_1, C_3\neq I$. Either $D_1$ or $D_3$ differs from the identity,
because Lemma~\ref{lemma:Asparse} shows that $HD_1$ and $D_3H$ cannot
both be equal to $H$. Futhermore, Lemma~\ref{lemma:Asparse} shows that
$HD_2H\neq I$, hence $D_2\neq I$. Therefore, we have at least six
gates.

Assume that $HC_1$ is not sparse.  Lemma~\ref{lemma:Anotsparse} shows
that $HD_1\neq H$ and $D_3H\neq H$ holds.  Therefore $D_1,D_3\neq I$.
Lemma~\ref{lemma:notsparse} shows that $HC_2H$ cannot be sparse, 
hence $C_2\neq I$.
Therefore, we have at least five gates. If $\det U\neq 1$, then 
$D_1,C_2,D_3$ cannot be the only nontrivial single qubit gates, 
as Lemma~\ref{lemma:obstruction} shows, 
proving that we have at least six elementary gates in that case.~\qed


\paragraph{More than Two Controlled-Not Gates.} 
Although it is undesirable to use more than two controlled-not gates,
we need to show (for mathematical completeness) that implementations
of a controlled-$U$ gate with three or more controlled-not gates
cannot reduce the total number of elementary gates.  Fortunately, it
turns out that the proof of this case is much simpler, because an
implementation with five or fewer gates can then have at most two
single qubit gates. Thus, the circuit can be expressed in the
following form:
\begin{equation}\label{circ:three}
\begin{emp}(50,50)
  setunit 2mm;
  qubits(2);
  
  gate(icnd 1, gpos 0, btex $U$ etex);
  label(btex $=$ etex,(QCxcoord+1/2QCstepsize, QCycoord[0]+3mm));
  QCxcoord := QCxcoord + QCstepsize;
  QCstepsize := QCstepsize-3mm;
  wires(2mm);
  circuit(1.0cm)(gpos 0,1, btex \small $P_1$ etex);
  gate(gpos 0, btex \small $A^{1\!-\!a}$ etex, 1, btex \small $A^a$ etex);
  circuit(1.0cm)(gpos 0,1, btex \small $P_2$ etex);
  gate(gpos 0, btex \small $B^{1\!-\!b}$ etex, 1, btex \small $B^b$ etex);
  circuit(1.0cm)(gpos 0,1, btex \small $P_3$ etex);
  wires(2mm);

\end{emp}
\end{equation}
where $a,b\in \{0,1\}$ determine the target bit of the single qubit
gates $A$ and~$B$, respectively.  The $P_i$ implement permutations of
the basis vectors realized by controlled-not operations. 

We collect some general observations about
implementations of controlled-$U$ gates with at most two single qubit
gates in Lemma~\ref{lemma:twosingles}--\ref{lemma:P2}.  It is then
shown in Lemma~\ref{lemma:threeC3}--\ref{lemma:threeC2} that an
implementation of a controlled-$U$ operation with three controlled-not
gates and at most two single qubit gates cannot exist, when $\tr U\neq
0$ and $\tr(UX)\neq 0$. The remaining cases are simple consequences of Lemma~\ref{lemma:twosingles}.

\begin{lemma}\label{lemma:twosingles} 
Let $U$ be a unitary $2\times 2$ matrix.  Suppose that there exists an
implementation of the controlled-$U$ gate with two single qubit gates
$A$ and $B$, and some controlled-not gates. Then $A$ is sparse if and
only if $B$ is sparse. 
\end{lemma}
\proof The input $\ket{00}$ remains unchanged by a controlled-$U$
operation. If $A$ is sparse, then the state after $P_2$ is of the form
$\alpha\ket{b_1b_0}$, with $b_1,b_0=0,1$. This state has to be mapped
by $B$ to a state of the form $\ket{c_1c_0}$, with $c_1,c_0=0,1$.
Thus, $B$ has to be sparse.  The same argument applied to the inverse
circuit shows that if $B$ is sparse, then $A$ has to be sparse as
well.~\qed

\begin{lemma}\label{lemma:crucial} 
Let $U$ be a unitary $2\times 2$ matrix.  
Suppose that there exists an
implementation of the controlled-$U$ gate with two single qubit gates
$A$ and $B$, and some controlled-not gates. If\/ $\tr U\neq 0$ and
$\tr(UX)\neq 0$, then $A$ and $B$ cannot be sparse.
\end{lemma}
\proof If $A$ and $B$ are sparse, then the circuit~(\ref{circ:three})
implements a monomial matrix. This would imply that $U$ is sparse,
contradicting either $\tr U\neq 0$ or $\tr(UX)\neq 0$.~\qed
\medskip

Denote by
$c(P_i)$ the number of controlled-not gates used to realize the permutation 
$P_i$ in (\ref{circ:three}). 
\begin{lemma}\label{lemma:P2}
Let $U$ be a unitary $2\times 2$ matrix. 
If\/ $\tr U\neq 0$ and $\tr(UX)\neq 0$, then $c(P_2)>0$ in the
circuit~(\ref{circ:three}).
\end{lemma}
\proof 
If $c(P_2)=0$, then (\ref{circ:three}) implies  
\begin{equation}
\begin{emp}(50,50)
  setunit 2mm;
  qubits(2);
  QCstepsize := QCstepsize-3mm;
  wires(2mm);
  circuit(1.0cm)(gpos 0,1, btex \small $P_1^\dagger$ etex);  
  gate(icnd 1, gpos 0, btex $U$ etex);
  circuit(1.0cm)(gpos 0,1, btex \small $P_3^\dagger$ etex);
  wires(2mm);
  label(btex $=$ etex,(QCxcoord+1/2QCstepsize, QCycoord[0]+3mm));
  QCxcoord := QCxcoord + QCstepsize;
  wires(2mm);
  gate(gpos 0, btex \small $A^{1\!-\!a}$ etex, 1, btex \small $A^a$ etex);
  gate(gpos 0, btex \small $B^{1\!-\!b}$ etex, 1, btex \small $B^b$ etex);
  wires(2mm);
\end{emp}
\end{equation}
The state input $\ket{00}$ will not be changed by the circuit on the
left hand side, because $P_1^\dagger$ and $P_3^\dagger$ are merely
sequences of controlled-not gates. On the other hand, if $a\neq b$,
then the circuit on the right hand side would map $\ket{00}$ to a
superposition of base states, because $A$ and $B$ are not sparse.
Thus, $a=b$, and $BA$ has to be a diagonal matrix. However, this would
imply that (\ref{circ:three}) is realizing a monomial matrix, which
contradicts $\tr U\neq 0$ or $\tr(UX)\neq 0$. Therefore, $c(P_2)$
cannot be zero.~\qed
\medskip

\paragraph{Three Controlled-Not Gates.}
We assume now that three controlled-not gates are used in the
circuit~(\ref{circ:three}), that is,  $c(P_1)+c(P_2)+c(P_3)=3$
and $0\le c(P_i)\le 3$.  We may assume without loss of generality that
$c(P_1)\ge c(P_3)$, for otherwise we can consider the inverse
circuits. Consequently, $c(P_3)$ is either $0$ or $1$. In
Lemma~\ref{lemma:threeC1} we consider the case $c(P_3)=1$, and in
Lemma~\ref{lemma:threeC2} the case $c(P_3)=0$.

\begin{lemma}\label{lemma:threeC3}
Let $U$ be a unitary $2\times 2$ matrix.
If\/ $\tr U\neq 0$ and $\tr(UX)\neq 0$, then $c(P_2)<3$ in the
circuit~(\ref{circ:three}) with three controlled-not gates.

\end{lemma}
\proof Seeking a contradiction, suppose that $c(P_2)=3$.
The three controlled-not operations in $P_2$ must have 
alternating target qubits, because otherwise it would be possible to
reduce the number of controlled-not gates.
Therefore, the circuit~(\ref{circ:three}) can be rewritten in the form:
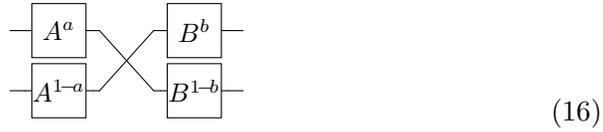
\begin{equation}
\begin{emp}(50,50)
  setunit 2mm;
  qubits(2);
  
  QCstepsize := QCstepsize-3mm;
 
  wires(2mm);
  gate(gpos 0, btex \small $A^{1\!-\!a}$ etex, 1, btex \small $A^a$ etex);
  perm(0,1);
  gate(gpos 0, btex \small $B^{1\!-\!b}$ etex, 1, btex \small $B^b$ etex); 
  wires(2mm);
\end{emp}
\end{equation}
Clearly, this circuit is not able to realize a controlled-$U$ operation,
because it cannot entangle any unentangled input state.~\qed

\begin{lemma}\label{lemma:threeC1}
Let $U$ be a unitary matrix with $\tr U\neq 0$ and $\tr(UX)\neq 0$.
If $c(P_3)=1$, then the circuit~(\ref{circ:three}) with three
controlled-not gates cannot implement a controlled-$U$ operation.
\end{lemma}
\proof We have $c(P_1)\ge c(P_3)$ and $c(P_2)>0$, hence $c(P_i)=1$ for $i=1,2,3$. The target bit of the controlled-not gate in $P_1$ (or $P_3$)
cannot be the least significant bit, because this would imply that
there exists an implementation of a controlled-$XU$ (or a
controlled-$UX$) gate with two controlled-not gates and at most four
elementary gates.  Proposition~\ref{proposition} shows that this is
not possible. Therefore, the circuit has to be of the form:
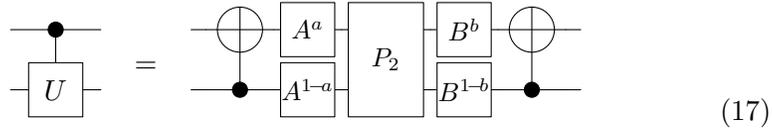
\begin{equation}\label{circ:threeup}
\begin{emp}(50,50)
  setunit 2mm;
  qubits(2);
  
  gate(icnd 1, gpos 0, btex $U$ etex);
  label(btex $=$ etex,(QCxcoord+1/2QCstepsize, QCycoord[0]+3mm));
  QCxcoord := QCxcoord + QCstepsize;
  QCstepsize := QCstepsize-3mm;
  wires(2mm);
  cnot(icnd 0, gpos 1);
  gate(gpos 0, btex \small $A^{1\!-\!a}$ etex, 1, btex \small $A^a$ etex);
  circuit(1.0cm)(gpos 0,1, btex \small $P_2$ etex);
  gate(gpos 0, btex \small $B^{1\!-\!b}$ etex, 1, btex \small $B^b$ etex);
  cnot(icnd 0, gpos 1);
  wires(2mm);

\end{emp}
\end{equation}
Regardless of the target bit of the controlled-not gate in $P_2$, we get
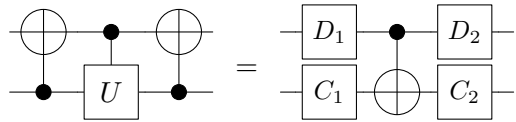
\begin{equation*}
\begin{emp}(50,50)
  setunit 2mm;
  qubits(2);
  QCstepsize := QCstepsize-3mm;
  cnot(icnd 0, gpos 1);  
  gate(icnd 1, gpos 0, btex $U$ etex);
  cnot(icnd 0, gpos 1);
  label(btex $=$ etex,(QCxcoord+1/2QCstepsize, QCycoord[0]+3mm));
  QCxcoord := QCxcoord + QCstepsize;

  wires(2mm);
  gate(gpos 0, btex \small $C_1$ etex, 1, btex \small $D_1$ etex);
  cnot(icnd 1, gpos 0);
  gate(gpos 0, btex \small $C_2$ etex, 1, btex \small $D_2$ etex);
  wires(2mm);

\end{emp}
\end{equation*}
for some unitary matrices $C_1, C_2, D_1, D_2$. If we input
$\ket{10}$, then the circuit on the left hand side produces an
entangled output state. This means $D_1$ has to be nonsparse, and
$C_1\ket{0}$ cannot be an eigenstate of $X$. The input $\ket{00}$ will
not produce an entangled state in the circuit on the left hand side,
but produces an entangled state on the right hand side.~\qed

\begin{lemma}\label{lemma:threeC2} 
Suppose that $U$ is a unitary $2\times 2$ matrix satisfying $\tr U\neq
0$ and $\tr UX\neq 0$. If $c(P_3)=0$
then the circuit~(\ref{circ:three}) with three controlled-not gates 
cannot implement a controlled-$U$ gate.
\end{lemma}
\proof 
Since $c(P_3)=0$ and $c(P_2)=1,2$, we have $c(P_1)=2,1$, respectively. 
Therefore, the circuit is of the form
\begin{equation}\label{circ:smallleft}
\begin{emp}(50,50)
  setunit 2mm;
  qubits(2);
  
  gate(icnd 1, gpos 0, btex $U$ etex);
  label(btex $=$ etex,(QCxcoord+1/2QCstepsize, QCycoord[0]+3mm));
  QCxcoord := QCxcoord + QCstepsize;
  QCstepsize := QCstepsize-3mm;
  wires(2mm);
  circuit(1.0cm)(gpos 0,1, btex \small $P_1$ etex);
  gate(gpos 0, btex \small $A^{1\!-\!a}$ etex, 1, btex \small $A^a$ etex);
  circuit(1.0cm)(gpos 0,1, btex \small $P_2$ etex);
  gate(gpos 0, btex \small $B^{1\!-\!b}$ etex, 1, btex \small $B^b$ etex);
  wires(2mm);

\end{emp}
\end{equation}
The permutation $P_1$ is of the form 
\begin{equation}\label{circ:P1form}
\begin{emp}(50,50)
  setunit 2mm;
  qubits(2);
  wires(2mm);
  circuit(1.0cm)(gpos 0,1, btex \small $P_1$ etex);
  wires(2mm);	
  label(btex $=$ etex,(QCxcoord+1/2QCstepsize, QCycoord[0]+4mm));
  QCxcoord := QCxcoord + QCstepsize;
  cnot(icnd 0, gpos 1);
  label(btex or etex,(QCxcoord+1/2QCstepsize, QCycoord[0]+4mm));
  QCxcoord := QCxcoord + QCstepsize;
  QCstepsize := QCstepsize - 3mm;
  cnot(icnd 0, gpos 1);
  cnot(icnd 1, gpos 0);
\end{emp}
\end{equation}
because otherwise it would be possible to realize a controlled-$XU$
operation with less than five operations, which contradicts
Proposition~\ref{proposition}. Notice that 
\begin{equation*}
\begin{emp}(50,50)
  setunit 2mm;
  qubits(2);
  
  QCstepsize := QCstepsize-3mm;
  
  wires(2mm);
  gate(gpos 1, btex $A^a$ etex, 0, btex $A^{1\!-\!a}$ etex);
  wires(2mm);
  label(btex $=$ etex,(QCxcoord+1/2QCstepsize, QCycoord[0]+4mm));
  QCxcoord := QCxcoord + QCstepsize;
  
  wires(2mm);
  perm(0,1);
  gate(gpos 1, btex $A^{1\!-\!a}$ etex, 0, btex $A^a$ etex);
  perm(0,1);
  wires(2mm);
\end{emp}
\end{equation*}
We will take advantage of this identity to derive the desired contradiction. 
Realizing that 
\begin{equation*}
\begin{emp}(50,50)
  setunit 2mm;
  qubits(2);
  QCstepsize := QCstepsize-3mm;  
  cnot(icnd 0, gpos 1);
  perm(0,1); 
  QCstepsize := QCstepsize+3mm;  
  label(btex $=$ etex,(QCxcoord+1/2QCstepsize, QCycoord[0]+4mm));
  QCxcoord := QCxcoord + QCstepsize;
  QCstepsize := QCstepsize-3mm;  

  cnot(icnd 1, gpos 0);
  cnot(icnd 0, gpos 1);
  QCstepsize := QCstepsize + 5mm;
  label(btex and etex,(QCxcoord+1/2QCstepsize, QCycoord[0]+4mm));
  QCxcoord := QCxcoord + QCstepsize;  
  QCstepsize := QCstepsize - 5mm;

  cnot(icnd 0, gpos 1);
  cnot(icnd 1, gpos 0);  
  perm(0,1); 
  QCstepsize := QCstepsize+3mm;  
  label(btex $=$ etex,(QCxcoord+1/2QCstepsize, QCycoord[0]+4mm));
  QCxcoord := QCxcoord + QCstepsize;
  QCstepsize := QCstepsize-3mm;  

  cnot(icnd 1, gpos 0);

\end{emp}
\end{equation*}
we find that the circuit (\ref{circ:smallleft}) can be re-written in the form 
\begin{equation}\label{circ:P2a}
\begin{emp}(50,50)
  setunit 2mm;
  qubits(2);
  
  gate(icnd 1, gpos 0, btex $U$ etex);
  label(btex $=$ etex,(QCxcoord+1/2QCstepsize, QCycoord[0]+3mm));
  QCxcoord := QCxcoord + QCstepsize;
  QCstepsize := QCstepsize-3mm;
  wires(2mm);
  cnot(icnd 1, gpos 0);
  cnot(icnd 0, gpos 1);
  gate(gpos 0, btex \small $A^{a}$ etex, 1, btex \small $A^{1\!-\!a}$ etex);
  perm(0,1);
  circuit(1.0cm)(gpos 0,1, btex \small $P_2$ etex);
  gate(gpos 0, btex \small $B^{1\!-\!b}$ etex, 1, btex \small $B^b$ etex);
  wires(2mm);

\end{emp}
\end{equation}  
or
\begin{equation}\label{circ:P2b}
\begin{emp}(50,50)
  setunit 2mm;
  qubits(2);
  
  gate(icnd 1, gpos 0, btex $U$ etex);
  label(btex $=$ etex,(QCxcoord+1/2QCstepsize, QCycoord[0]+3mm));
  QCxcoord := QCxcoord + QCstepsize;
  QCstepsize := QCstepsize-3mm;
  wires(2mm);
  cnot(icnd 1, gpos 0);
  gate(gpos 0, btex \small $A^{a}$ etex, 1, btex \small $A^{1\!-\!a}$ etex);
  perm(0,1);
  circuit(1.0cm)(gpos 0,1, btex \small $P_2$ etex);
  gate(gpos 0, btex \small $B^{1\!-\!b}$ etex, 1, btex \small $B^b$ etex);
  wires(2mm);

\end{emp}
\end{equation} 
depending on the form of $P_1$ shown in (\ref{circ:P1form}), respectively.  
The circuits (\ref{circ:P2a}) and (\ref{circ:P2b}) can both be
simplified to contain at most five elementary gates, by reducing the
combination of the swap operation with $P_2$ to merely 1 and 2
controlled-not gates, respectively.  This would imply that the
controlled-$XU$ gate can be realized with at most four elementary
gates, contradicting Proposition~\ref{proposition}.~\qed

\begin{proposition}
Let $U$ be a unitary matrix with $\tr U\neq 0$ and $\tr(UX)\neq 0$.
It is impossible to implement a controlled-$U$ gate with two or fewer
single qubit gates and three controlled-not gates.
\end{proposition}
\proof This follows from Lemma~\ref{lemma:threeC1}--\ref{lemma:threeC2}.~\qed

\paragraph{Four or more Controlled-Not Gates.}
If we have more than three controlled-not gates, then an
implementation with less than six elementary gates is not possible,
because of Lemma~\ref{lemma:crucial}.
\medskip

\paragraph{Summary.} We have shown that a generic controlled-$U$ gate cannot be
implemented with less than six elementary gates. In fact, we could
rule out implementations based on a single controlled-not
gate. Implementations with two controlled-not gates are possible, but
at least four single qubit gates are necessary. The previous
discussion showed that this gate count cannot be improved by
implementations based on three or more controlled-not gates. This
concludes the proof of Theorem~A.~\qed

\section{Further Ramifications}\label{sec:nongeneric}
Let $m(U)$ denote the minimal number of elementary gates that are
needed to implement a controlled-$U$ gate.  We know
from~\cite{bbc-egqc-95} that $m(U)\le 6$.  We have shown in the
previous section that $m(U)=6$ provided that $U$ is generic.  We will
show next that $m(U)\le 5$ when $U$ is not generic.
\bigskip

\goodbreak
\noindent\textbf{Theorem B} \textit{
Let $U$ be a unitary $2\times 2$ matrix. Let $\phi$ and $\phi_0$ be real numbers in the range $0<\phi<2\pi$ and $0\le\phi_0<2\pi$.\\[-4ex] 
\begin{tabbing}aa \=  \kill
a) \> If\/ $U=I$, then $m(U)=0$.\\
b) \>If\/ $U=e^{i\phi}I$, then $m(U)=1$.\\
c) \>If\/ $U=X$, then $m(U)=1$.\\
d) \>If\/ $U=e^{i\phi}X$, then $m(U)=2$. \\
e) \>If\/ $U=e^{i\phi_0}Z$, then $m(U)=3$.\\
f) \>If\/ $\tr U=0$, $\det U=-1$, $U\neq \pm X$, then $m(U)=3$.\\
g) \>If\/ $\tr U=0$, $\det U\neq -1$, $U\neq e^{i\phi_0}X$, 
$U\neq e^{i\phi_0}Z$, then $m(U)=4$. \\ 
h) \>If\/ $\tr UX=0$, $\tr U\neq 0$, $\det U=1$, $U\neq \pm I$, then $m(U)=4$. \\
i) \>If\/ $\tr UX=0$, $\tr U\neq 0$, $\det U\neq 1$, $U\neq e^{i\phi_0}I$, then $m(U)=5$. \\
j) \>If\/ $\det U=1$, $\tr U\neq 0$, $\tr UX\neq 0$, then $m(U)=5$.  
\end{tabbing}} 
\medskip

\noindent Theorem B captures all non-generic cases. The upper bounds on the number
of gates are straightforward to see with the help of
Table~\ref{table}.
\medskip

\begin{table}[htb]
$$
\begin{array}{l|l|ll}
 \mbox{Case} & \mbox{Form} & \mbox{Circuit} \\
\hline 
& & & \\[-1.5ex]
\raisebox{0.6cm}{\mbox{if }}\,  \raisebox{0.6cm}{\mbox{$\tr U=0$}} & \raisebox{0.6cm}{\mbox{$U=e^{i\phi}P XP^\dagger$}}  &
\begin{emp}(0,0)
  setunit 2mm;
  qubits(2);
  QCstepsize := QCstepsize -3mm;
  wires(2mm);
  gate(gpos 1, btex $E$ etex, 0, btex \small $P^\dagger $ etex);
  cnot(icnd 1, gpos 0);
  gate(gpos 0, btex \small $P$ etex);
  wires(2mm);
\end{emp} 
& \raisebox{0.6cm}{\mbox{(C1)}} \\
\raisebox{0.6cm}{\mbox{else if }}\, \raisebox{0.6cm}{\mbox{$\tr(UX)=0$}} &  \raisebox{0.6cm}{\mbox{$U=e^{i\phi}P XP^\dagger X$}} & 
\begin{emp}(50,50)
  setunit 2mm;
  qubits(2);
  
  QCstepsize := QCstepsize-4mm;
  wires(2mm);
  cnot(icnd 1, gpos 0);
  gate(gpos 1, btex $E$ etex, 0, btex \small $P^\dagger$ etex);
  cnot(icnd 1, gpos 0);
  gate(gpos 0, btex \small $P$ etex);
  wires(2mm);
\end{emp}
& \raisebox{0.6cm}{\mbox{(C2)}} \\
\raisebox{0.6cm}{\mbox{else if }}\, \raisebox{0.6cm}{\mbox{$\det U=1$}} &  \raisebox{0.6cm}{\mbox{$U=CXBXA$}}&
\begin{emp}(50,50)
  setunit 2mm;
  qubits(2);
  
  QCstepsize := QCstepsize-4mm;
  wires(2mm);
  gate(gpos 0, btex $A$ etex);
  cnot(icnd 1, gpos 0);
  gate(gpos 0, btex $B$ etex);
  cnot(icnd 1, gpos 0);
  gate(gpos 0, btex $C$ etex);
  wires(2mm);
\end{emp}
& \raisebox{0.6cm}{\mbox{(C3)}} \\
\raisebox{0.6cm}{\mbox{else }} & \raisebox{0.6cm}{\mbox{$U=e^{i\phi} CXBXA$}} &  \begin{emp}(50,50)
  setunit 2mm;
  qubits(2);
  
  QCstepsize := QCstepsize-4mm;
  wires(2mm);
  gate(gpos 1, btex $E$ etex, 0, btex $A$ etex);
  cnot(icnd 1, gpos 0);
  gate(gpos 0, btex $B$ etex);
  cnot(icnd 1, gpos 0);
  gate(gpos 0, btex $C$ etex);
  wires(2mm);
\end{emp}
& \raisebox{0.6cm}{\mbox{(C4)}} 
\end{array}
$$
\caption{Quantum circuits for the implementation of controlled-$U$ gates.}\label{table}
\end{table}

We formally prove tight upper bounds on $m(U)$ in the following simple
Lemma.
\begin{lemma}\label{lemma:upperbounds}
The number  $m(U)$ of elementary gates given in the statement of
Theorem~B are sufficient to realize the corresponding controlled-$U$
gates.
\end{lemma}
\proof 
The cases $a)$--$c)$ of Theorem B are obvious. 
\smallskip

\noindent\textit{Case d)} If $U=e^{i\phi}X$, then the circuit (C1) in Table~\ref{table} 
with $P=I$ and
$E=\diag(1,e^{i\phi})$ implements a controlled-$U$ gate, hence
$m(U)\le 2$.
\smallskip

\noindent\textit{Case e)} If $U=e^{i\phi_0}Z$, then 
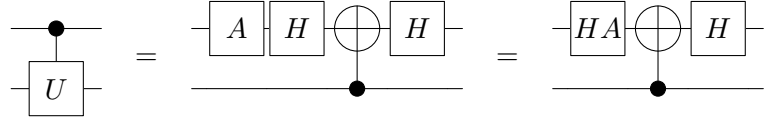
\begin{center}
\begin{emp}(50,50)
  setunit 2mm;
  qubits(2);
  
  gate(icnd 1, gpos 0, btex $U$ etex);
  label(btex $=$ etex,(QCxcoord+1/2QCstepsize, QCycoord[0]+4mm));
  QCxcoord := QCxcoord + QCstepsize;

  QCstepsize := QCstepsize-4mm;
  wires(2mm);
  gate(gpos 1, btex $A$ etex);
  gate(gpos 1, btex $H$ etex);
  cnot(icnd 0, gpos 1);
  gate(gpos 1, btex $H$ etex);
  wires(2mm);

  QCstepsize := QCstepsize+4mm;
  label(btex $=$ etex,(QCxcoord+1/2QCstepsize, QCycoord[0]+4mm));
  QCxcoord := QCxcoord + QCstepsize;
  QCstepsize := QCstepsize-4mm;

  wires(2mm);
  gate(gpos 1, btex $HA$ etex);
  cnot(icnd 0, gpos 1);
  gate(gpos 1, btex $H$ etex);
  wires(2mm);
\end{emp}
\end{center}
with $A=\diag(1,e^{i\phi_0})$, hence
$m(U)\le 3$.
\smallskip

\noindent\textit{Cases f, g)} If $\tr U=0$, then $U$ is of the form
$U=e^{i\phi}PXP^\dagger$. Therefore, circuit (C1) of Table~\ref{table}
with $E=\diag(1,e^{i\phi})$ shows that $m(U)\le 4$, which proves the
upper bound of $g)$.  If in addition $\det U=-1$, then $E=I$ in (C1),
whence $m(U)\le 3$.
\smallskip

\noindent\textit{Cases h, i)} If $\tr(UX)=0$, then the matrix $U$ is of
the form $U=e^{i\phi}PXP^\dagger X$ for some unitary matrix $P$, and
$\phi\in \R$. Let $E=\diag(1,e^{i\phi})$.  The circuit (C2) in
Table~\ref{table} shows that $m(U)\le 5$, proving the upper bound of $i)$. If in addition
$\det U=1$, then necessarily $E=I$, hence $m(U)\le 4$, proving the
upper bound of $h)$.
\smallskip 

\noindent Case $j)$. If $\det U=1$, then it is possible to realize the
controlled-$U$ gate in the form (C3), as was shown in \cite{bbc-egqc-95}, hence
$m(U)\le 5$.~\qed
\medskip

It remains to prove the lower bounds on $m(U)$. 
The following lemma allows to prove the cases a)--f):
\begin{lemma}\label{lemma:I}
Suppose that $U$ is a unitary $2\times 2$ matrix.  If the controlled-$U$
gate can be implemented with one single qubit gate $A$ and some
controlled-not gates, then $U$ has to be of the form $U=e^{i\phi}I$ or
$U=e^{i\phi}X$ for some $\phi\in \R$. Furthermore, $A$ has to be diagonal and can be assumed to be of 
the form $A=\diag(1,e^{i\phi})$.
\end{lemma}
\proof The controlled-$U$ gate is realized by a circuit of the form
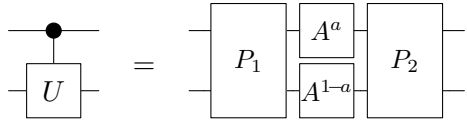
\begin{equation*}
\begin{emp}(50,50)
  setunit 2mm;
  qubits(2);
  
  gate(icnd 1, gpos 0, btex $U$ etex);
  label(btex $=$ etex,(QCxcoord+1/2QCstepsize, QCycoord[0]+3mm));
  QCxcoord := QCxcoord + QCstepsize;
  QCstepsize := QCstepsize-3mm;
  wires(2mm);
  circuit(1.0cm)(gpos 0,1, btex \small $P_1$ etex);
  gate(gpos 0, btex \small $A^{1\!-\!a}$ etex, 1, btex \small $A^a$ etex);
  circuit(1.0cm)(gpos 0,1, btex \small $P_2$ etex);
  wires(2mm);

\end{emp}
\end{equation*}
where $P_1$ and $P_2$ are permutations realized by controlled-not
operations, and the target bit of $A$ is selected by $a\in \{0,1\}$.
The state $\ket{00}$ remains invariant under the action of a
controlled-$U$ gate. Since $P_1\ket{00}=\ket{00}=P_2^\dagger\ket{00}$,
it follows that $(A^a\otimes A^{1-a})\ket{00}=\ket{00}$, whence $A$
has to be a diagonal matrix. By multiplying with an irrelevant global
phase factor, we can assume that $A$ is of the form
$A=\diag(1,e^{i\phi})$.  Notice that the phase of exactly two out of
the four computational base states are changed to $e^{i\phi}$ by $A$,
hence $U$ has to be of the stated form.~\qed

\begin{lemma} Parts a)--f) of Theorem B hold. 
\end{lemma}
\proof It is clear that no gate is needed to implement a
controlled-identity gate, whence $a)$ holds. In $b)$, only one phase
gate is needed to affect the phase change, hence $b)$ is true.  At
least one controlled-not gate is needed in the cases \textit{c)--f)},
because $U$ has two different eigenvalues $a$ and $-a$.  We have
$m(U)\ge 2$ in cases \textit{d)--f)}, because $U\neq X$. If
$U=e^{i\phi}X$, then $m(U)=2$ by Lemma~\ref{lemma:upperbounds}, which
proves $d)$.

We have $U=e^{i\phi_0}Z$ in case \textit{e)}. This gate can affect a
phase change, hence at least one single qubit gate is needed.
Lemma~\ref{lemma:I} shows that circuits with one single qubit gate and
controlled-not gate cannot implement $U=e^{i\phi_0}Z$. Therefore,
another single qubit gate is needed, that is, $m(e^{i\phi_0}Z)\ge 3$,
whence $m(e^{i\phi_0}Z)=3$ by Lemma~\ref{lemma:upperbounds}.

In case $f)$, $U$ is a unitary matrix with $\tr U=0$, $\det U=-1$, and
$U\neq \pm X$. We know that $m(U)\ge 2$. Two controlled-not gates
cannot implement such a gate because of the determinant condition.
Lemma~\ref{lemma:I} shows that a single qubit gate and a
controlled-not gate cannot implement~$U$. Therefore, $m(U)\ge 3$,
hence $m(U)=3$ by Lemma~\ref{lemma:upperbounds}.~\qed
\medskip

The remaining cases need a little bit more work. The next lemma gives
some partial information about circuit with two single qubit gates and
some controlled-not gates. Lemma~\ref{lemma:twosingles} showed that
$A$ and $B$ are either both sparse or both non-sparse. The sparse case is covered by the following lemma:

\begin{lemma}\label{lemma:II}
Suppose that $U$ is a unitary $2\times 2$ matrix. 
If the controlled-$U$ gate can be implemented with 
two sparse single-qubit gates $A$ and $B$, and some controlled-not gates,  
then the matrix $U$ has to be of the form 
$U = e^{i\phi}I$, $U = e^{i\phi}X$, 
$U = \diag(e^{i\phi}, e^{-i\phi})$, or
$U = \antidiag(e^{i\phi}, e^{-i\phi})$, where $\phi\in \R$. 
\end{lemma}
\proof Since the matrices 
$A$ and $B$ are sparse, $U$ has to be sparse as well, that is,
$U=\diag(e^{i\phi_1},e^{i\phi_2})$ or
$U=\antidiag(e^{i\phi_1},e^{i\phi_2})$.

Suppose that $A$ or $B$ is of the form $e^{i\phi}I$, $\phi\in
\R$. Such a gate affects only a global phase change, and thus may be
deleted. It follows from Lemma~\ref{lemma:I} that $U$ is of the
desired form.

Suppose that $A$ and $B$ are not multiples of the identity matrix. We
may multiply $A$ and $B$ with global phase factors without changing
the functionality of the circuit.  Therefore, we can assume that $A$
and $B$ are of the form $A=X^a\diag(1,e^{i\phi_A})$ and
$B=\diag(1,e^{i\phi_B})X^b$ for some $a, b\in \{0,1\}$, and
$0<\phi_A,\phi_B<2\pi$. Consider the four computational base states
$\mathcal{B}=\{\ket{00},\ket{01},\ket{10},\ket{11}\}$ as inputs to our
circuit.  The circuit realizes a monomial matrix, because $A$ and $B$
are sparse.  Therefore, ignoring order, the output of these four input
states is given by a set of four states
$\{\alpha\ket{00},\beta\ket{01},\gamma\ket{10},\delta\ket{11}\}$ with
phase factors $\alpha, \beta, \gamma,\delta$.  Since the circuit
realizes a controlled-$U$ operation with sparse $U$, the multiset of
these four phase factors should be of the form
$P_U=\{1,1,e^{i\phi_1},e^{i\phi_2}\}$. 

Notice that $A$ and $B$ each affect a phase change in exactly two of
the four computational base states. During the evolution of a input
state from $\mathcal{B}$, the state might be multiplied by phases
$e^{i\phi_A}$ and $e^{i\phi_B}$. If we record the combinatorial
possibilities, then the multiset of phase factors
$\{\alpha,\beta,\gamma,\delta\}$ can be of the form:
\begin{tabbing} aaa\=\kill
a)\> $\{ e^{i\phi_A}, e^{i\phi_A}, e^{i\phi_B}, e^{i\phi_B}\},$ provided
$A$ and $B$ affect disjoint inputs,\\[1ex]
b)\> $\{ 1, e^{i\phi_A}, e^{i\phi_B}, e^{i(\phi_A+\phi_B)}\},$ provided
a single input is affected by $A$ and $B$,\\[1ex] 
c)\> $\{ 1, 1, e^{i(\phi_A+\phi_B)}, e^{i(\phi_A+\phi_B)}\},$ provided 
$A, B$ affect the same two inputs. 
\end{tabbing}
We compare these multisets with $P_U$ to derive some constraints about $U$. 
It is clear that case~a) cannot occur, because $e^{i\phi_A},
e^{i\phi_B}\neq 1$. In case~b), we necessarily have 
$\phi_A=-\phi_B$, therefore $U$ is of the form
$U=\diag(e^{i\phi_A},e^{-i\phi_A})$ or
$U=\antidiag(e^{i\phi_A},e^{-i\phi_A})$.  In case~c), it follows that
$U$ is of the form $U=e^{i\phi}I$ or $U=e^{i\phi}X$ with
$\phi=\phi_A+\phi_B$.~\qed
\medskip

\begin{remark}\label{lemma:III}
Suppose that a unitary matrix $U$ is of the form 
$U = e^{i\phi}X$, $U = e^{i\phi}I$,
$U = \diag(e^{i\phi}, e^{-i\phi})$, or 
$U = \antidiag(e^{i\phi}, e^{-i\phi})$. Notice that 
\vspace*{-1.5ex}
\begin{tabbing}aaa\=\kill
i)\> if\/ $\tr(U) = 0$,  then $U = e^{i\phi}X$, $U = \pm iZ$, or 
$U = \antidiag(e^{i\phi}, e^{-i\phi})$,\\
ii)\> if\/ $\tr(U)\neq 0$, then 
$U = e^{i\phi}I$, or
$U = \diag(e^{i\phi}, e^{-i\phi})$.
\end{tabbing}
\end{remark}

The following lemma proves case \textit{g)} of Theorem B: 

\begin{lemma}\label{lemma:V}
Suppose that $U$ is a unitary $2\times 2$ matrix. 
If $\tr(U) = 0$, $\det(U) \neq -1$, $U \neq e^{i\phi}X$, 
and $U \neq e^{i\phi}Z$, then $m(U)=4$. 
\end{lemma}
\proof Seeking a contradiction, we assume that $m(U)\le 3$.  The
condition $\tr U=0$ implies that $U\neq e^{i\phi}I$, 
i.e., at least one controlled-not gate is
needed in the implementation.

Suppose that at least two controlled-not gates are used in the
implementation. This means that at most one single qubit gate can be
used. Lemma~\ref{lemma:I} shows that $U$ would have to be of the form
$U=e^{i\phi}X$, contradicting the assumptions. Therefore, the
potential implementation of $U$ must have one controlled-not gate.

Suppose now that one controlled-not gate and at most two single qubit
gates $A$ and $B$ are used in the implementation. The matrices $A$ and
$B$ are either both sparse or both not sparse by
Lemma~\ref{lemma:twosingles}. 
\medskip

\noindent\textit{Case 1.} 
Assume that $A$ and $B$ are sparse. Then $U$ has to be
sparse. According to Lemma~\ref{lemma:II} and Remark~\ref{lemma:III}
the matrix $U$ would have to be of the form $U=e^{i\phi}X$, $U=\pm iZ$
or $U=\antidiag(e^{i\phi},e^{-i\phi})$.  None of these matrices
satisfies the assumptions of the lemma, contradiction.
\medskip

\noindent\textit{Case 2.}  Suppose now that $A$ and $B$ are not
sparse.  If the controlled-not gate has the same target bit as the
controlled-$U$ gate, then the circuit is of the form (\ref{circ:single}). 
Lemma~\ref{lemma:single} shows that $A_1$ and $A_2$ are sparse. 
It follows that the circuit has to be of the form
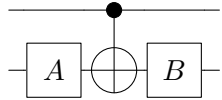
\begin{equation*}
\begin{emp}(50,50)
  setunit 2mm;
  qubits(2);

  QCstepsize := QCstepsize - 4mm;
  wires(2mm);
  gate(gpos 0, btex $A$ etex);
  cnot(icnd 1, gpos 0);
  gate(gpos 0, btex $B$ etex);
  wires(2mm);
\end{emp}
\end{equation*}
This implies $BA=e^{i\phi}I$ and $BXA=e^{i\phi}U$, whence $\det U=-1$,
contradicting the assumptions.

Assume now that target bit of the controlled-not gate is the most
significant bit. One easily sees that the two single qubit gates have
to act on the most significant bit as well, i.e., the circuit is of the form 
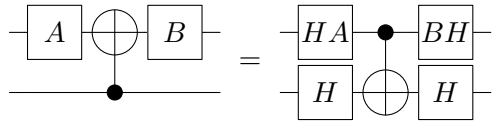
\begin{equation*}
\begin{emp}(50,50)
  setunit 2mm;
  qubits(2);

  QCstepsize := QCstepsize - 4mm;
  wires(2mm);
  gate(gpos 1, btex $A$ etex);
  cnot(icnd 0, gpos 1);
  gate(gpos 1, btex $B$ etex);
  wires(2mm);

  label(btex $=$ etex,(QCxcoord+1/2QCstepsize, QCycoord[0]+4mm));
  QCxcoord := QCxcoord + QCstepsize;
  
  wires(2mm);
  gate(gpos 0, btex $H$ etex, 1, btex $HA$ etex);
  cnot(icnd 1, gpos 0);
  gate(gpos 0, btex $H$ etex, 1, btex $BH$ etex);
  wires(2mm);

\end{emp}
\end{equation*}
It follows from Lemma~\ref{lemma:single} that $HA$ and $BH$ are
sparse, whence $A$ and $B$ are both not sparse.  Notice that
$\ket{0}$, $\ket{1}$ are eigenvectors of $U$, say with eigenvalues
$\alpha_0$, $\alpha_1$, respectively. 
Corollary~\ref{corollary} shows that
$BA=\diag(1,\alpha_0)$, and $BXA=\diag(1,\alpha_1)$.
Comparing determinants shows that
$\alpha_2=-\alpha_1$. This implies that $U$ is of the form 
$U=\diag(\alpha_1,\alpha_2)=\alpha_1Z$, contradicting the assumptions.

Therefore, we can conclude that $m(U)\ge 4$. We obtain $m(U)=4$ with Lemma~\ref{lemma:upperbounds}.~\qed 
\medskip

We proceed with the proof of case h) of Theorem B. 
\begin{lemma}
Let $U$ be a unitary $2\times 2$ matrix.
If $\tr(UX)=0$, $\tr U\neq 0$, $\det U=1$, $U\neq \pm I$, then $m(U)=4$. 
\end{lemma}
\proof Seeking a contradiction, we assume that $m(U)\le 3$.
Lemma~\ref{lemma:onecontrol} shows that more than one controlled-not
gate has to be used in the implementation of the controlled-$U$ gate.
We cannot have an implementation with three controlled-not gates,
because the matrix corresponding to this circuit would have
determinant $-1$. In the remaining case, one single qubit gate and two
controlled-not gates are used for the
implementation. Lemma~\ref{lemma:I} shows that a solution $U$ with
$\det U=1$ would have to be of the form $U=\pm I$ or $U=-X$, all of
which contradict the assumptions.  We can conclude that $m(U)\ge 4$, hence $m(U)=4$ by Lemma~\ref{lemma:upperbounds}.~\qed
\medskip

The next lemma covers case \textit{i)} of Theorem B. 
\begin{lemma}\label{lemma:VII}
Suppose that $U$ is a unitary $2\times 2$ matrix satisfying $\tr(U)
\neq 0$, $\det U\neq 1$, $U\neq e^{i\phi}I$, and $\tr(UX) = 0$. Then
five elementary gates are necessary and sufficient to implement such a
controlled-$U$ gate.
\end{lemma}
\proof Seeking a contradiction, we assume that $m(U)\le 4$. 
\medskip

\noindent\textit{Case 1.} Suppose that the implementation uses at
least \textit{three} controlled-not gates. According to
Lemma~\ref{lemma:I}, $U$ would have to be of the form $U=e^{i\phi}I$
or $U=e^{i\phi}X$, which contradicts the assumptions $U\neq e^{i\phi}$
and $\tr(U)\neq 0$.
\medskip

\noindent\textit{Case 2.} Suppose that only one controlled-not gate is
used in the implementation of the controlled-$U$ gate. It follows that
$\tr U=0$ by Lemma~\ref{lemma:onecontrol}.  This contradicts the
assumption $\tr U\neq 0$.
\medskip

\noindent\textit{Case 3.}  Suppose that two controlled-not gates and at
most two single qubit gates $A$ and $B$ are used in the implementation
of the controlled-$U$ gate. We know from Lemma~\ref{lemma:twosingles} 
that $A$ and $B$ are either both sparse or both not sparse. 
\medskip

\noindent\textit{Case 3.1.} 
Suppose that $A$ and $B$ are both sparse. It follows from  
Lemma~\ref{lemma:II} and
Remark~\ref{lemma:III} that $U$ has to be of the form
$U=e^{i\phi}I$ or $U=\diag(e^{i\phi},e^{-i\phi})$, which contradicts the assumptions $U\neq e^{i\phi}I$ and $\det U\neq 1$.

\noindent\textit{Case 3.2.} Suppose that $A$ and $B$ are both not
sparse. We distinguish four different cases, depending on the target
bit of the two controlled-not gates.
\medskip

\textit{Case} $\downarrow \downarrow$. Suppose that the target bit of both
controlled-not gates is the least significant bit. It follows from
Lemma~\ref{lemma:notsparse} and Lemma~\ref{lemma:Asparse} that the
single qubit gates have to act on the least significant bit as well. 
Assume without loss of generality that the control-$U$ operation is implemented by a circuit of the form 
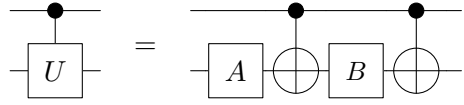
\begin{center}
\begin{emp}(50,50)
  setunit 2mm;
  qubits(2);
  
  gate(icnd 1, gpos 0, btex $U$ etex);
  label(btex $=$ etex,(QCxcoord+1/2QCstepsize, QCycoord[0]+3mm));
  QCxcoord := QCxcoord + QCstepsize;
  QCstepsize := QCstepsize-4mm;
  wires(2mm);
  gate(gpos 0, btex $A$ etex);
  cnot(icnd 1, gpos 0);
  gate(gpos 0, btex \small $B$ etex);
  cnot(icnd 1, gpos 0);
  wires(2mm);
\end{emp}
\end{center}
Comparing the result of the input $\ket{0}\otimes \ket{\psi}$ and
$\ket{1}\otimes \ket{\psi}$ shows that $e^{i\theta}I = BA$ and 
$e^{i\theta}U = XBXA$. Comparing determinants yields $\det U=1$, which
contradicts our assumptions.
\medskip

\textit{Case} $\uparrow \uparrow$. Suppose that the most significant
bit is the target bit of both controlled-not gates. There must be a
single qubit gate, say $B$, on the most significant bit between the
two controlled-not gates, because of Lemma~\ref{lemma:notsparse}.  The
other single quantum bit gate has to be on the most significant bit as
well, in order to map $\ket{00}$ to $\ket{00}$.

Therefore, we may assume without loss of generality that the circuit is of the form 
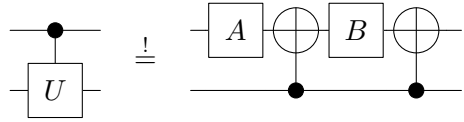
\begin{center}
\begin{emp}(50,50)
  setunit 2mm;
  qubits(2);

  gate(icnd 1, gpos 0, btex $U$ etex);
  label.top(btex $\stackrel{!}{=}$ etex,(QCxcoord+1/2QCstepsize, QCycoord[0]+2mm));
  QCxcoord := QCxcoord + QCstepsize;
  QCstepsize := QCstepsize-4mm;   

  wires(2mm);
  gate(gpos 1, btex $A$ etex);
  cnot(icnd 0, gpos 1);
  gate(gpos 1, btex $B$ etex);
  cnot(icnd 0, gpos 1);
  wires(2mm);
\end{emp}
\end{center}
Notice that
$\ket{0}$, $\ket{1}$ are eigenvectors of $U$, say with eigenvalues
$\alpha_0$, $\alpha_1$, respectively. 
Corollary~\ref{corollary} shows that
$BA=\diag(1,\alpha_0)$, and $XBXA=\diag(1,\alpha_1)$.
Comparing determinants shows that
$\alpha_0=\alpha_1$. This implies that $U$ is of the form 
$U=\diag(\alpha_0,\alpha_1)=\alpha_0I$, contradicting the assumptions.
\medskip

\textit{Cases} $\uparrow\downarrow$ and $\downarrow \uparrow$. Finally, consider the
case that the two controlled-not gates have different target bits.  We
may assume that the first controlled-not gate has the most significant
bit as  target bit. If this is not the case, the we simply consider
the inverse circuit. The circuit is of the general form 
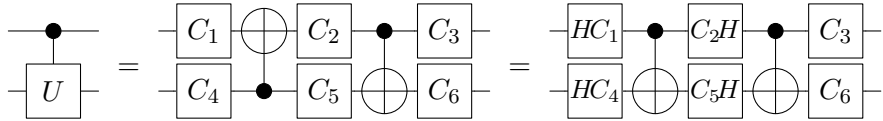
\begin{equation*}
\begin{emp}(50,50)
  setunit 2mm;
  qubits(2);
  gate(icnd 1, gpos 0, btex $U$ etex);  
  QCstepsize := QCstepsize-4mm;

  label(btex $=$ etex,(QCxcoord+1/2QCstepsize, QCycoord[0]+3mm));
  QCxcoord := QCxcoord + QCstepsize;
  
  wires(2mm);
  gate(gpos 1, btex $C_1$ etex, 0, btex $C_4$ etex);
  cnot(icnd 0, gpos 1);
  gate(gpos 1, btex $C_2$ etex, 0, btex $C_5$ etex);
  cnot(icnd 1, gpos 0);
  gate(gpos 1, btex $C_3$ etex, 0, btex $C_6$ etex);
  wires(2mm);
  label(btex $=$ etex,(QCxcoord+1/2QCstepsize, QCycoord[0]+3mm));
  QCxcoord := QCxcoord + QCstepsize;

  wires(2mm);
  gate(gpos 1, btex\small $H\!C_1$ etex, 0, btex \small $H\!C_4$ etex);
  cnot(icnd 1, gpos 0);
  gate(gpos 1, btex\small $C_2\!H$ etex, 0, btex \small $C_5\!H$ etex);
  cnot(icnd 1, gpos 0);
  gate(gpos 1, btex $C_3$ etex, 0, btex  $C_6$ etex);
  wires(2mm);
\end{emp}
\end{equation*}
where two of the matrices $C_i$ are given by $A$ and $B$, and the
remaining four are identity matrices. We use the circuit on the right
hand side to derive a contradiction.

Lemma~\ref{lemma:Asparse} and Lemma~\ref{lemma:eigenvector} show that
$C_5H\neq H$, hence $C_5\neq I$. Consequently, at least one of the
matrices $C_1$ or $C_2$ has to be the identity matrix. Therefore,
$HC_1$ or $C_2H$ has to be nonsparse, whence $C_3$ is nonsparse by
Lemma~\ref{lemma:notsparse}. It follows that $C_1=C_2=C_4=C_6=I$.
However, we know from Lemma~\ref{lemma:double} that $HC_1$ and $C_2H$
cannot both be equal to $H$, thus it is impossible that $C_1=C_2=I$,
contradiction.

Therefore, we can conclude that it is impossible to implement a
controlled-$U$ gate with $m(U)\le 4$ operations. It follows from Lemma~\ref{lemma:upperbounds} that $m(U)=5$, which concludes the proof.~\qed
\medskip

It remains to show case \textit{j)} of Theorem B: 

\begin{lemma}
Let $U$ be a unitary $2\times 2$ matrix. If $\det U=1$, $\tr U\neq 0$,
and $\tr(UX)\neq 0$, then $m(U)=5$.
\end{lemma}
\proof It is not possible to implement such a controlled-$U$ gate with
only one controlled-not gate, cf.~Lemma~\ref{lemma:onecontrol}. If two
controlled-not gates are used in the implementation, then
Proposition~\ref{proposition} shows that three additional single qubit
gates are necessary. 

Assume that $m(U)\le 4$ elementary gates are enough. If three or more
controlled-not gates are used in the implementation, then at most one
single qubit gate can be used. Lemma~\ref{lemma:I} shows that $U$
would have to be of the form $U=e^{i\phi}I$ or $U=e^{i\phi}X$,
contradicting $\tr(UX)\neq 0$ and $\tr U\neq 0$.

Therefore, $m(U)\ge 5$. It was shown in \cite{bbc-egqc-95} that
$m(U)\le 5$ when $\det U=1$, which proves the claim.~\qed

\section{Conclusions}
We have derived the minimal number of elementary gates that are
necessary in any implementation of a controlled unitary gate.  It
would be interesting to know tight lower bounds for other fundamental
constructions of quantum circuits.  In particular, it would be nice to
know the minimal number of elementary gates that are needed to realize
doubly controlled-$U$ gates, such as the Toffoli gate.  
\medskip

\noindent\textbf{Acknowledgments.}  We thank Martin R\"otteler for
numerous comments that helped to improve this paper.

\end{empfile}

\renewcommand{\refname}{Reference}

\end{document}